\documentclass[
  ,draft            
  ]
  {aipproc}

\layoutstyle{8x11single}

\def\la{\mathrel{\hbox{\rlap{\hbox{\lower4pt\hbox{$\sim$}}}{\raise2pt\hbox{$<$}}}}}
\def\ga{\mathrel{\hbox{\rlap{\hbox{\lower4pt\hbox{$\sim$}}}{\raise2pt\hbox{$>$}}}}}

\begin{document}

\title{Observations of ultraluminous X-ray sources}


\author{Jeanette C. Gladstone}{
  address={Department of Physics, University of Alberta, Edmonton, Alberta, T6G 2G7, Canada}
}

\begin{abstract}

Ultraluminous X-ray sources (ULXs), first observed $\sim$30 years ago, have been argued as extreme stellar mass black hole binaries or a new class intermediate mass black hole. In order to settle this debate, scientists have utilised a wide range of telescopes, exploiting large sections of the electro-magnetic spectrum. Here we review some of the insight gained from these observational studies, collating an overview of our current position in ULX research.

\end{abstract}

\maketitle


\section{Introduction}

Ultraluminous X-ray Sources (ULXs) are point like, extragalactic, non-nuclear objects with inferred X-ray luminosities in excess of the Eddington limit ($L_{Edd} \simeq$10$^{39}$ erg s$^{-1}$) for a stellar mass black hole (StMBH, 3 -- 20 M$_\odot$), radiating isotropically. The simplest way to explain this is to assume a larger mass compact object, with proposed masses ranging from 10$^2$ -- 10$^{4}$M$_\odot$ (e.g. Colberts \& Mushotzky 1999), accreting at sub-Eddington rates. Black holes of this size provide an attractive opportunity to explain the `missing link' in the mass scale of black holes, and present possible formation routes for super-massive black holes (SMBHs) and the formation of galaxies (e.g. Madau \& Rees 2001; Ebisuzaki et al. 2001). 

An alternate theory is that we are instead observing StMBHs\footnote{Here we define StMBHs to be the product of the end point of a single stellar evolution ($\la$ 100$M_\odot$), and use this throughout the article.} that {\it appear} to be emitting above the Eddington limit, circumventing the Eddington limit via relativistic or geometric beaming. Relativistic beaming occurs as a consequence of the observer looking down the funnel of a jet (K{\"o}rding et al. 2002), resulting in Doppler boosted emission. This occurs in either the low/hard (LHS) or (occasionally) very high state (VHS), when jet emission is present. Geometric beaming occurs at higher mass accretion rates as the disc approaches the Eddington limit and when radiation pressure becomes significant (e.g. King et al. 2001; King 2008; 2009). The inflowing material, being fed through the accretion disc reaches the Eddington limit (locally) at a point known as the spherization or trapping radius. To keep the local radiation energy release close to Eddington, the excess energy and matter is lost via a strong outflow. This collimation causes a beaming effect that boosts the apparent emission from the source for an observer along the cone of emission.  

The third option is that we are in fact observing super-Eddington accretion onto a StMBH, forcing a change in the accretion geometry. There are many suggestions for possible geometries at such high accretion rates. One is the slim disc model, where an increase in mass accretion rate causes the disc to become so optically thick that photons are advected radially leading to additional cooling (Abramowicz et al. 1988). An alternative to this was proposed by Poutanen et al. (2007), in which it is predicted that we would observe a cool accretion disc and a photosphere (radiating a blackbody-like spectrum) emitting a cool ($\sim$ 1 keV) thermal component  along with a possible outflow/wind. 

Above we have outlined some of the current theories proposed to explain the emission from ULXs. Throughout this paper we will discuss multi-wavelength observations of these sources, focussing on the insights that can be obtained with regard to  the underlying compact object, in an attempt to unlock the nature of these systems. 

\section{Early ULX surveys}

The first catalogues of ULXs were created using {\it ROSAT} (e.g. Colbert \& Mushotzky 1999; Roberts \& Warwick 2000; Colbert \& Ptak 2002). These surveys showed that non-nuclear luminous X-ray sources appeared in $\la$ 20 per cent the galaxies sampled (Miller \& Colbert 2004) and covered a wider range of luminosities ($\la$ a few $\times$ 10$^{40}$ erg s$^{-1}$). 

Although these sources are rare, over-densities have been observed in some galaxies such as the Antennae, or the Cartwheel. In a survey of the Antennae, it was found that $\ga$ 10 sources emitting in the ultraluminous regime (e.g. Fabbiano et al. 2003). The mass of these objects was initially estimated by assuming they were X-ray binary systems containing a black hole emitting at the Eddington limit. The luminosity of these sources suggested that we were observing black holes of mass $M_{\rm BH} \ga$ 10 -- (few $\times$) 100 $M_\odot$ (assuming no beaming; Miller \& Colbert 2004), however their accretion rate is unknown.

\section{Optical studies}

Starting from the largest size scales and moving inwards, ULXs have been found preferentially in actively star-forming galaxies (Kaaret 2005), along with starburst, interacting/merging and dwarf galaxies (Swartz et al. 2008) preferably with low metallicities (Pakull \& Mirioni 2002; Zampieri \& Roberts 2009; Mapelli et al. 2009), in the range $\frac{1}{6}  Z_\odot  <  Z  <  \frac{1}{2}  Z_\odot$ (Ripamonti et al 2010). Such low metallicities are important in the formation of large StMBHs. The low metallicity reduces the loss of material via a wind during a stars life, enhancing the maximum possible black hole mass (e.g. Mapelli et al 2010), it also enhances the probability of forming High Mass X-ray Binaries (HMXBs; Linden et al 2010).

\begin{figure}
  \includegraphics[height=0.21\textheight]{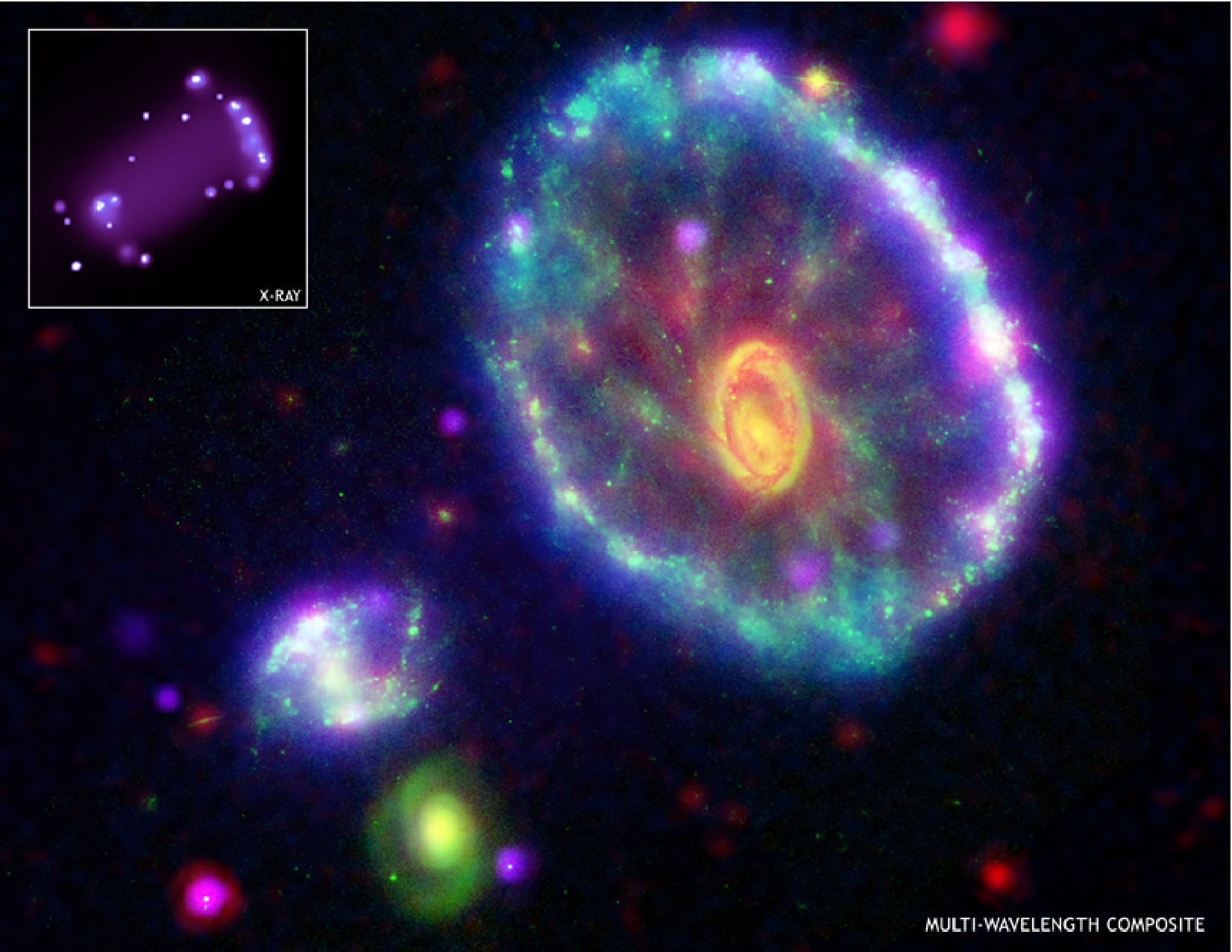}
 \includegraphics[height=0.31\textheight, angle=90]{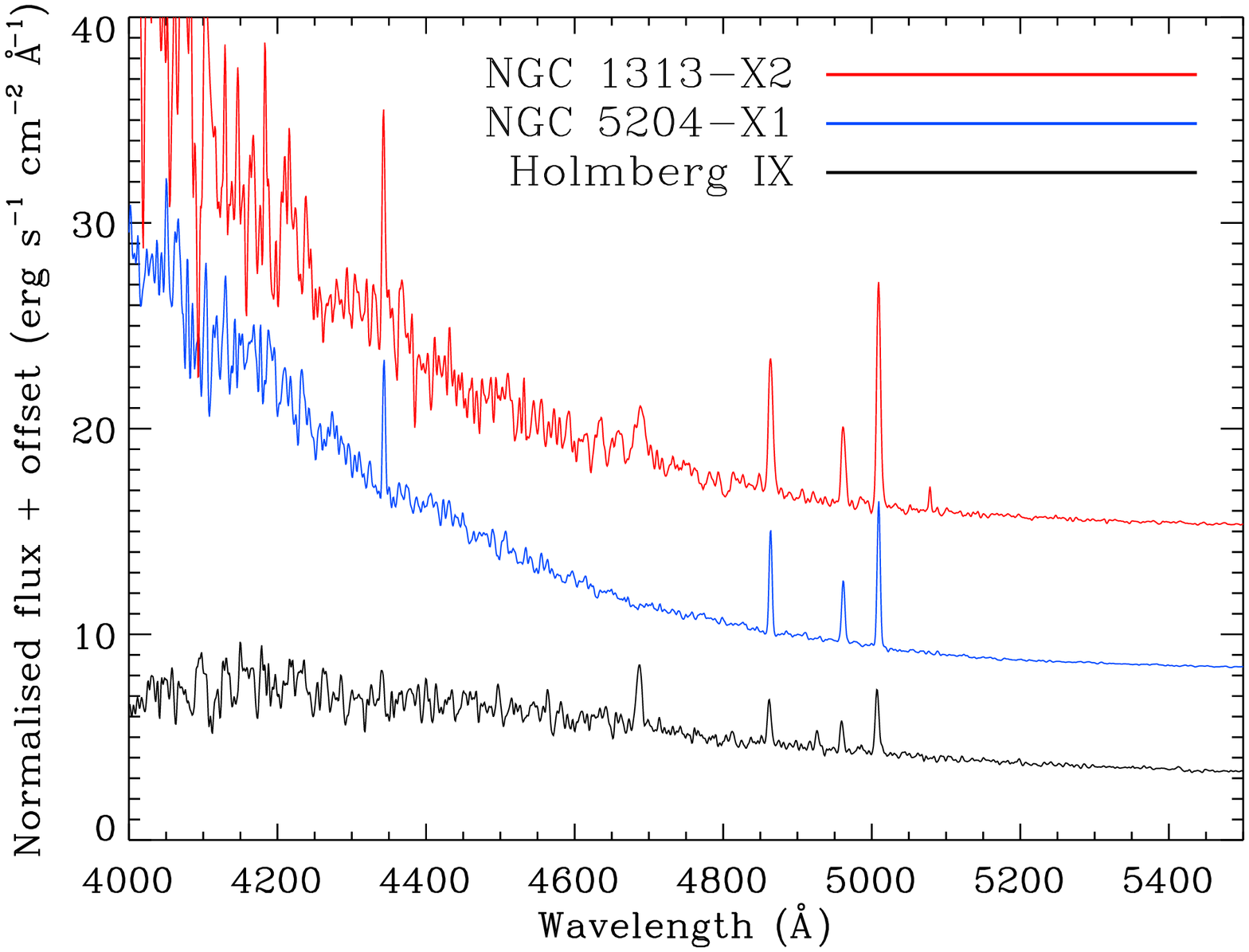}
  \caption{({\it Left}) Cartwheel galaxy: purple - {\it Chandra} Observatory; ultraviolet/blue - Galaxy Evolution Explorer satellite; visible/green - Hubble Space Telescope; infrared/red - Spitzer Space Telescope. The insert highlights the large X-ray population contained within this galaxy (Image composite: NASA/JPL/Caltech/P.Appleton et al. X-ray: NASA/CXC/A.Wolter \& G.Trinchieri et al.). ({\it Right}) Pilot spectra of thee ULXs taken by the Gemini Observatory (Roberts et al. 2010), we note the presence of strong nebula emission lines in all cases and the He \textsc{ii} 4686 {\AA} line in two cases (thought to be associated with the accretion disc). }
   \label{fig:cartwheel}
\end{figure}

An observed correlation between the number of ULXs and the global star-formation rate in spirals has also been noted (Swartz et al. 2004; Liu et al. 2006). The sheer number of sources observed in starburst galaxies together with their short lifetimes (implied by their location in regions of active star formation) means that, if these systems were all to contain IMBHs, an unrealistically large underlying population must be present (see Figure \ref{fig:cartwheel} ({\it Left}); King 2004). Instead it is much more likely that the bulk of the population are the most extreme examples of HMXBs, that somehow exceed or circumvent the Eddington limit. These high mass companions provide a natural origin for the high mass transfer rates required to power the observed luminosities (Rappaport et al. 2005). 

An association with star-forming regions (Fabbiano, Zezas \& Murray 2001; Lira et al. 2002; Gao et al 2003) and the unbroken luminosity function connecting ULXs to the standard X-ray binary population (Grimm, Gilfanov \& Sunyaev 2003) also indicates that we may be observing HMXB systems.  

\subsubsection{ULX nebulae}

Pakull \& Mirioni (2002) presented a study of ULX environments in which some ULXs appear to reside in emission nebulae of a few hundred parsecs in diameter. They reported evidence of both low and high ionisation emission lines within their spectra, indicating that they could be powered by either shock or photo-ionised radiation. If these lines are signatures of photo-ionisation, these features can be used to place limits on the X-ray emission powering it. By doing this Pakull \& Mirioni (2002) concluded that the optical emission was consistent with an isotropic X-ray emitter. This would indicate the presence of an IMBH at its centre. 

Observation of other ULX nebulae (e.g. Ho II X-1; Kaaret et al. 2004) showed evidence of X-ray ionisation, with high levels of excitation near the ULX. Despite the higher impact of X-ray emission, these sources also show a lack of beamed emission, suggesting that we are observing either IMBHs or some form of super-Eddington accretion. 

Further analysis of some systems revealed evidence of cavities in the nebulae of some systems, possibly cleared by shocks (e.g. NGC 5204 X-1; Roberts et al. 2002), while explorations of their spectra showed evidence of both photoionisation and shocks (e.g. Pakull \& Mirioni 2003). Also, some ULX nebulae show signatures of strong winds and/or jets (Abolmasov et al. 2007). The presence of cavities/bubbles and winds/jets suggests a link to features observed around both StMBHs and SMBHs. Such comparisons can give us insights into the feedback that occurs within these systems (Pakull et al. 2010) and into the strength of outflows (Russell et al 2010).  Although this work is still in the early stages of development, results indicate that ULXs could be a {\it heterogeneous} sample of sources.

\subsubsection{Stellar counterparts}

The search for optical counterparts has revealed a number of faint blue potential counterparts ($m_{\rm V} \sim$ 22 -- 26; Roberts et al. 2008), suggesting O or B type stars. We must also consider that the emission could be contaminated by reprocessed X-rays from the accretion disc/stellar surface (e.g. Copperwheat et al. 2005; Madhusudhan et al. 2008). Models created to account of such X-ray irradiation can estimate the underlying stellar spectral type and the mass of the black hole, with the aide of colour-magnitude diagrams. However, studies using such models have provided conflicting values for black hole mass, Madhusudhan et al. (2008) indicated a slight preference for IMBHs, whilst another shows support for StMBHs (Copperwheat et al. 2007). In either case, the emission was impacted greatly by X-ray irradiation (increase of $\sim$ 0.5 -- 5 magnitudes, depending on assumptions; Copperwheat et al. 2007). Such a dramatic effect must be taken into account in future studies of these systems. 

Studies of the stellar environment of ULXs have also provided insight into the nature of the companion. For example, the colour-magnitude diagram of the neighbouring stars of NGC 1313 X-2 show that the companion is $\sim$ 20 Myrs old with an upper mass limit of 12 $M_\odot$ (Grise et a 2008). This confirms that these systems are HMXBs, but it does not provide  the mass of the black hole. An {\it HST} photometric study of the same object revealed a possible $\sim$ 6 day orbital period (Liu et al. 2009, although further analysis is required to confirm this). When combined with the results of Grise et al (2008), Patruno \& Zampieri (2010) indicated that $M_{BH} \simeq$ 50 -- 100 $M_\odot$, suggesting a StMBH residing in a HMXB system. 

Spectroscopic studies of these systems have also recently been undertaken. Such studies are difficult due to the luminosity of these systems, but are possible with some of the worlds largest telescopes (e.g. Gemini, VLT; for an example of data received, see Figure \ref{fig:cartwheel} ({\it Right}). By obtaining a series of observations, it may be possible to obtain radial velocity curve measurements from emission/absorption features associated with either the accretion disc or the companion star. This provides a direct mass measurement based on constraints placed by the binary orbit. Only a small number of optical pilot spectra have been taken and published to date. Roberts et al. (2001) studied the counterpart of NGC 5204 X-1, finding a blue, featureless spectrum. Similar results have been found in NGC 1313 X-2 (Zampieri et al. 2004).  Pilot spectra of other sources have revealed a lack of spectral features, and slopes that appear non-stellar in origin (see Figure \ref{fig:cartwheel} ({\it Right}); Roberts et al. 2010; Gladstone et al. {\it in prep}). 

Radial velocity measurements are beginning, however, initial results from a survey of two ULXs (NGC 1313 X-2 \& Ho IX X-1) found that radial velocity variations are observed, but that they may not be sinusoidal (Roberts et al. 2010; Gladstone et al. {\it in prep}), however further analysis is required to confirm this. The analysis of these spectra also indicated that the He \textsc{ii}  emission line, which is associated with the accretion disc, appears to be highly variable (Pakull et al. 2006; Roberts et al. 2010). Such dramatic variation on short timescales implies that, at a minimum, the He \textsc{ii} emission originates within 24 light-hours of the ULX (Roberts et al. 2010). Motch et al. (2010) reported on studies of ULX P13, in NGC 7793, stating that variations in the He \textsc{ii} line are present, and superimposed on a photospheric spectrum. This reveals the possible presence of a late B type supergiant companion of between 10 -- 20 $M_\odot$. Radial velocity variations were also detected for this source, but again further data and analysis is required to confirm its period and nature. 

\section{Radio Emission}

Radio emission is observed regularly from sources residing in the LHS, and occasionally from those in the VHS (although the type of radio emission would vary depending on state). By combining this information with the luminosity of these systems, it would indicate the presence of an IMBH if these sources reside in the LHS (steady state emission), or a large StMBH -- IMBH residing in the VHS (more varied emission). 

To date, very few radio detection have been made, but  steady state emission has been observed in two of the ULXs within M82, and in one of those located in NGC 4736 (K{\"o}rding et al. 2005). In this case it would suggest the LHS therefore the radio and X-ray emission are related as follows $L_{\rm X} \propto L_{\rm R}^{1.38} M^{0.81}$ (Falcke et al. 2004), implying a mass of $M_{\rm BH} \simeq 10^2 - 10^5 M_\odot$ (although there may be some source confusion, further observations are required to confirm this). 

A study by Miller et al. (2005) identified region of diffuse resolved radio emission in the area surrounding Ho II X-1, coincident with an extended H \textsc{ii} region. Their investigation indicates that this could not be powered by beamed emission, and so would need to be powered by either some form of super-Eddington accretion onto a StMBH or by sub-Eddington accretion onto an IMBH. However, the nebula surrounding NGC 5408 X-1 also shows evidence of radio emission (Kaaret et al. 2003), although the authors note that this emission could be the result of relativistic beaming from a jet. Later work revealed this emission to be slightly extended, with a diameter of 35 -- 46 pc (Lang et al. 2007). Their results could not rule out relativistically beamed emission, but highlighted the fact that that the nature of the emission appeared to be comparable with SS 433. This would indicate that we are observing radio emission powered by outflows from a super-Eddington system, similar to the W50 nebula around SS 433 (Dubner et al. 1998)

\section{X-ray analysis}

The study of ULXs began with observations in the X-ray band, and since then there have been a numerous observations. Studies have focused on both the spectra and timing for these systems, as we search for insights into their nature. These are discussed in more detail below, however, for a more detailed discussion of X-ray analysis of ULXs (pre-2007) we refer the reader to Roberts (2007) and references therein. 

\subsection{X-ray spectral studies}

Early spectral studies, with the current generation of X-ray telescopes were fit with a standard canonical disc plus power-law continuum model, a simple model that has been used extensively with Galactic X-ray binary systems to describe accretion in standard states. The resultant ULX fits indicated the presence of a remarkably cool disc (soft excess; kT $\sim$ 150 eV) temperature, implying a $M_{BH} \sim$ few $\times 10^{4} M_\odot$ (e.g Miller et al. 2003). However, recent results indicated the presence of a spectral break above $\sim$ 3 keV (Stobbart et al. 2006; Gladstone et al. 2009), a feature that is not present in the standard accretion states, and so should not be present if we were observing IMBHs. Re-evaluation of some of the highest quality ULX spectra also revealed that they could be equally well explained by models that infer extreme accretion rates onto a StMBH (e.g. Stobbart et al. 2006, Goncalves \& Soria 2006) and found a potentially new state - the {\it ultraluminous state} in which a source exhibit the presence of both a soft excess and a break at higher energies (in the 0.3 -- 10 keV band pass; Gladstone et al. 2009).

Gladstone et al. (2011) discussed some of the current physical models for ULXs; the slim disc, disc plus Comptonisation, the energetically coupled disc-corona and the reflection dominated model. A simplified slim disc models has been applied to ULX spectra on a number of occasions (e.g. Vierdayanti et al. 2006). However, when applied to the highest quality data, Gladstone et al. (2009) finds it an inadequate descriptor. It is proposed, therefore, that a more complex version of the slim disc (e.g. Sadowski et al. 2010) should instead be applied to investigate this theory further. 

\begin{figure}
  \includegraphics[height=0.19\textheight, angle=0]{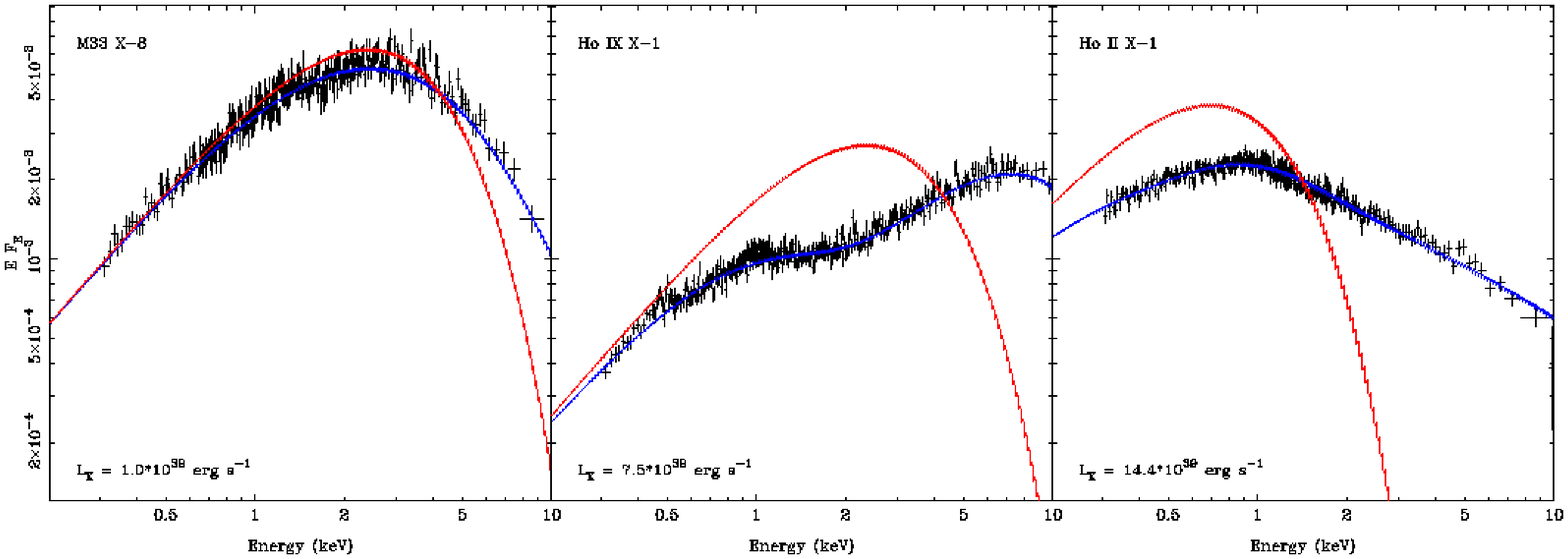}
  \includegraphics[height=0.19\textheight]{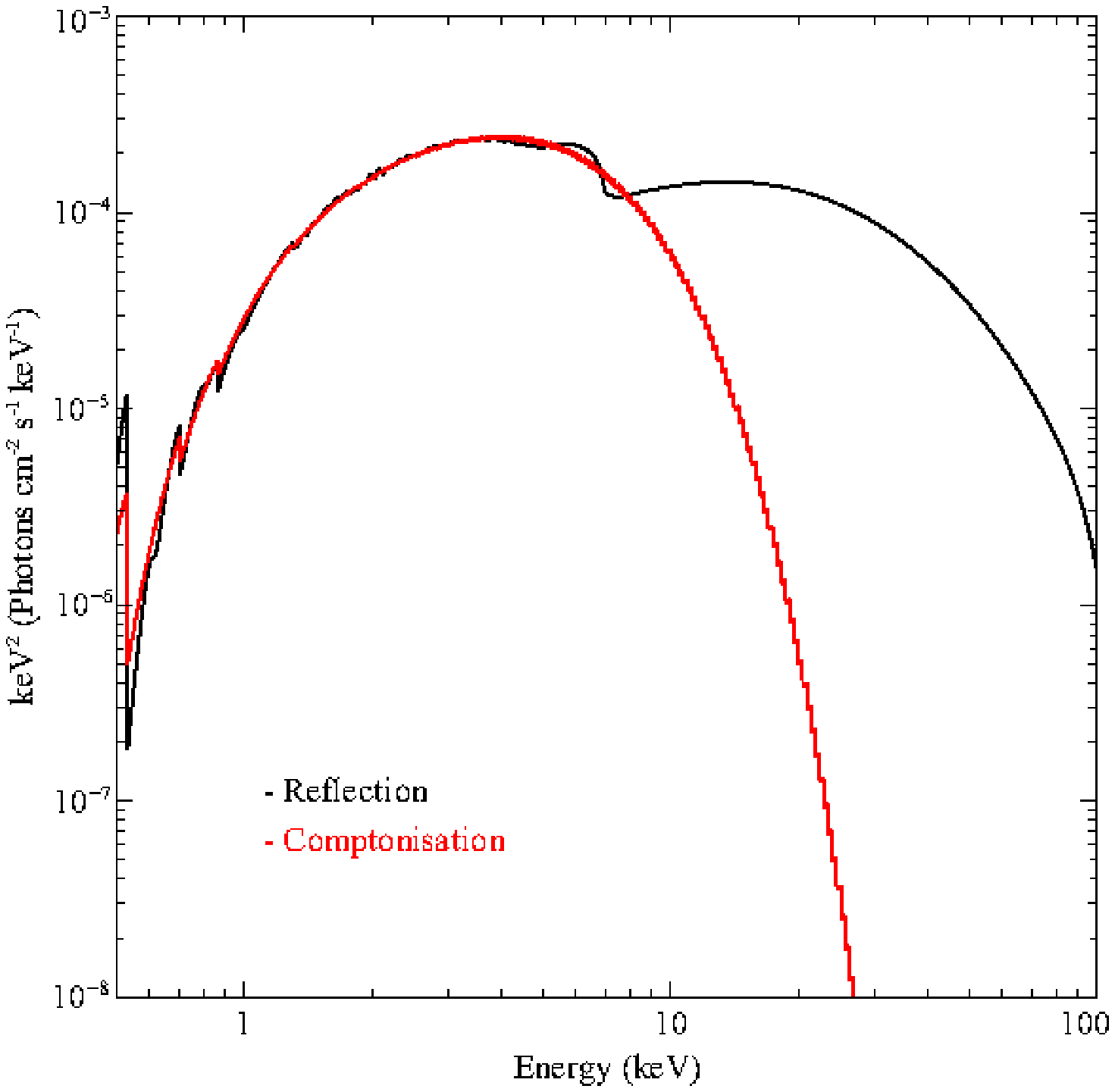}
  \caption{({\it Left}) {\it XMM-Newton} EPIC-pn X-ray spectra for three ULXs from Gladstone et al. (2009). The spectra are de-absorbed \& unfolded and show one spectrum from each suggested spectral regime: the high (M33 X-8); ultraluminous (Ho IX X-1); and extreme ultraluminous (possibly photosphere-dominated; Ho II X-1) states. The intrinsic disc spectrum, recovered when the corona is accounted for, is shown in red (see Gladstone et al. 2009 for more details). ({\it Right}) Comparison of the ultraluminous state model and the X-ray reflection model extended to 100 keV. The two short similarities below 10 keV, but the difference in flux level above 10 keV covers many orders of magnitude (see Walton et al. 2011 for more details).}
  \label{fig:ul state}
\end{figure}

Comptonisation models revealed a preference for a cool disc with an optically thick corona ($\tau >$ 8, Stobbert et al. 2006; $\tau >$ 6, Gladstone et al. 2009). This is very different from standard black hole systems (LHS - $\tau \la$ 2, $kT_e \sim$ 50 keV; VHS - $\tau \sim$ 3, $kT_e \sim$ 20 keV), indicating that ULXs may be residing in a state that is more extreme than even the VHS, and that we may be observing super-Eddington accretion. However, the discs remain cool, which could also be used as support the argument for IMBHs (Gladstone et al. 2009). The model assumed that an optically thick corona does not intercept the line of sight to inner regions of the accretion disc, and that the underlying disc spectrum is independent of the presence of a corona (Kubota \& Done 2004). These flawed assumptions can be overcome by applying the energetically coupled disc plus corona model to the sample. Resulting fits indicate that there are three possible regimes (see Figure \ref{fig:ul state} ({\it Left})); the high/VHS or some form of high mass accretion rate modified disc-dominated state, the ultraluminous state and the extreme ultraluminous state (where we see possible evidence for a photosphere and/or outflow (e.g Begelman, King \& Pringle 2006; Poutanen et al 2007). 

An alternative scenario is the X-ray reflection model (Caballero-Garc{\'i}a \& Fabian 2010). The authors chose to apply this model for similar reasons to those outlined above, reflected iron lines have been observed in the spectrum of both StMBHs (e.g. Miller 2007) \& SMBHs (e.g. Tanaka et al. 1995). In this case the authors propose that ULX spectra can be explained by an X-ray irradiated disc around a rapidly spinning black hole. Results indicated that ULXs fall into two major groups, a reflection dominated or power-law dominated state. It is argued that these correspond to two of the regimes discussed in Miniutti \& Fabian (2004), in which height of the primary source of X-rays is varied. This allows the observer to see differing amounts of the continuum and  reflection components.

Walton et al. (2011) also compared two of these leading (energetically coupled disc-corona \& X-ray reflection) models. The authors found  that the fits are statistically similar with moderate quality data, highlighting the need for a wider band-pass to differentiate between models (see Figure \ref{fig:ul state} ({\it Right})). The ultraluminous state model continues falling off in flux, whilst the reflection model continues its power-law slope out to larger energies. The ability to look to higher energies would also settle the debate over the mass of these sources, by clearly demonstrating the spectral state of these systems. Although we still cannot confirm statistically the structure of the accretion geometry, it seems that the majority of the current physical models indicate the presence of StMBHs when using the highest quality spectral data.

\subsubsection{X-ray spectral variability}

Spectral variability of StMBHs have proved invaluable in understanding of accretion states (e.g. McClintock \& Remillard 2006), yet it is an avenue that remains under-explored in ULXs. Some simple surveys have been performed, revealing that the majority of ULXs display a general hardening as the luminosity of the system increases (e.g. Antennae population, Fabbiano et al. 2003; Homberg II X-1, Dewangan et al. in 2004; NGC 5204 X-1, Roberts et al. 2006). This is the opposite of what is seen in the evolution of the standard accretion states, this could support the idea of a new accretion state. However, recent studies have highlighted a heterogeneity in ULXs. A study of the ULX population of NGC 4485 \& 4490 (Gladstone \& Roberts 2009), along with a survey of some of the most observed ULXs (Kajava \& Poutanen 2009), revealed that not only a $L_X$ vs. $\Gamma$ correlation, but also an inverse correlation in some sources, with Gladstone \& Roberts (2009) noting the change in behaviour coincided with a possible change in spectral shape (and possible change in state). 

Kajava \& Poutanen (2009) also noted that the cool disc (soft excess) component varied as $L \propto T^{-3.5}$. This implies that the inner most radius of the disc moves out and the luminosity increases, which is strikingly different from the standard decrease in inner radius observed in Galactic StMBH systems ($L \propto T^4$). This could be due to material and energy being lost at these radii as a result of launching a wind/outflow.

Recent work attempted to test the findings of Gladstone et al. (2009) by studying a series of reasonable quality spectra of NGC 1313 X-2 (Pintore \& Zampieri 2010) and Ho IX X-1 (Vierdayanti et al. 2010). Pintore \& Zampieri (2010) presented analysis of moderate quality {\it XMM-Newton} observations, while Vierdayanti et al. (2010) presented {\it XMM-Newton} and {\it Swift} data (grouped by count rate) to study spectral variations over long timescales. When the data is fit with a combined disc and Comptonising corona model, the authors found that the spectra are well fit by a cool optically thick corona, which showed a decrease in the coronal temperature and increasing optical depth as the luminosity of the system increases. This appears to be in agreement with the predictions of Gladstone et al. (2009). However, when Kong et al. (2010) fit grouped {\it Swift} data of Ho IX X-1 in a similar way to Vierdayanti et al. (2010), they did not see the same trends. They found that all the spectra are well fit by a duel thermal model, which they argued indicates super-Eddington accretion. Clearly further variability studies are required to disentangle the spectral evolution of these systems, and provide further insights into the accretion geometry of these systems.

\subsection{X-ray temporal studies}

Long-term temporal variability studies have suffered, in part, due to ULXs being too faint to be picked up by monitoring missions such as {\it RXTE} (with the exception of M82 X-1, Kaaret et al. 2006), and so can only be observed during pointed observations. This leads to large gaps in the light curves, making it difficult to track the trends of these systems. Some authors have managed to find evidence for variation (up to $\sim$ an order of magnitude) on time-scales of months to years, with the use of multi-mission data. Examples include the ULXs in M81 (La Parola et al. 2001), Ho II (Miyaji et al 2001), NGC 4485 \& 4490 (Roberts et al. 2002a; Gladstone \& Roberts 2009) and M51 (Terashima \& Wilson 2003). If we consider shorter time-scales, of days to weeks, we find that fewer studies have been reported. Roberts et al. (2006) presented a study of NGC 5204 X-1 in which factor 5 variations were observed over a time-scales of days. However, no sign of variability were observed over timescales of hours or less. 

Intra-observational analysis for both StMBHs and SMBHs is explored using power spectral densities (PSD). It has been shown that many features are common to both StMBHs and SMBHs (McHardy et al. 2006; McClintock \& Remillard 2006; van der Klis 2006). However, ULXs seem to have suppressed variability (Heil et al. 2009), with only a small number of features observed. These include a break at 2.5 mHz in the PSD of NGC 5408 X-1 (Soria et al. 2004) indicating a mass of $\sim$100 $M_\odot$ (see Heil et al. 2009 for more detailed discussions and analysis).

Quasi-periodic oscillations (QPOs) have also been detected in only three cases; M82 X-1 (Strohmayer \& Mushotzky 2003), another source in M82 (Feng et al. 2010), \& NGC 5408 X-1 (Strohmayer et al. 2007; 2010). These QPOs has been used as evidence for IMBHs, by using black hole mass - QPO frequency scaling. Recent re-analysis of the data of NGC 5408 X-1 has shown a different interpretation, however, Middleton et al. (2010) has shown that the spectra, PSD and variability of this source is better match to a model of super-Eddington accretion than that of the LHS. This would indicate that the QPO present in this source is more analogous to the mHz QPO observed in GRS 1915+105, indicating the presence of a StMBH.

\section{The nature of these systems?}

Since the advent of {\it XMM-Newton} \& the {\it Chandra} Space telescopes, the study of ULXs had advanced dramatically. Initial results from these telescopes indicated the presence of massive black holes, more massive than could be explained by the end point of (single) stellar evolution (implying IMBHs). Early X-ray emission provided tentative support, but the strongest X-ray evidence for IMBHs to date comes from the observations of hyper-luminous X-ray source (HLXs). These are ULXs emitting at $\ga 10^{41}$ erg s$^{-1}$. The brightest of these sources is ESO 243-49 HLX-1 (Farrell et al. 2009; 2010) with a lower mass limit of $\sim$ 500 $M_\odot$. Optical observations of the nebulae associated with some ULX systems also lent support to IMBHs, by indicating that the X-ray emission powering these systems was not strongly beamed. There have been few Radio observations of these source, with many only revealing limits on possible emission. In the few instances where Radio emission has been observed, scaling relations indicate that these ULXs contain IMBHs. The presence of QPOs in three ULX has been pointed to as some of the strongest evidence for the presence of IMBHs in ULX systems. 

More recent studies in many of these wavebands have revealed a different scenario. Studies of the X-ray spectra of these systems has revealed the presence of a break at $\sim$ 3 -- 7 keV, a feature not present in the LHS, while the leading physical models used to describe these systems indicate that we are observing super-Eddington accretion flows. X-ray timing studies have also revealed suppressed variability in many sources, again a feature not seen in the standard accretion states. This combined with the recent re-analysis of one QPO source, which indicates a better match to super-Eddington theory than the LHS, suggest that we are in fact observing some of the most extreme StMBH X-ray binary systems in the Universe. It is not just X-ray analysis that supports this theory. Optical photometry studies reveal the presence of high mass companion stars, and an association with young, low metallicity, star forming regions (regions in which large StMBHs are thought to form). The over-abundance in some galaxies has also been used as evidence to support the presence of StMBHs. 

Although there appears to be mounting evidence for extreme StMBHs powering ULXs, we must note that each of these mass estimates are degenerate, relying on assumption to gain mass estimates. To settle the debate, we must instead turn to more direct estimates of the mass, and for this we need to gain dynamical mass estimates of these systems, a challenging task that is still in the early stages at present, but may yet reveal the underlying nature of these systems.



\begin{theacknowledgments}

JCG would like to thank the organisers of this meeting for inviting her to present this review. She would also like to thank her collaborators for some of the work presented here, as well as colleagues and researchers in this and related fields for their discussions and guidance. JCG would also like to apologise to those authors whose work is not mentioned here, due to the lack of space and the wealth of informationavailable on these fascinating sources. 
  
\end{theacknowledgments}



\bibliographystyle{aipproc}   




\end{document}